\documentclass[twocolumn,showpacs,preprintnumbers,amsmath,amssymb,floatfix,superscriptaddress]{revtex4-1}

\usepackage[dvips]{graphicx}% Include figure files
\usepackage{dcolumn}% Align table columns on decimal point
\usepackage{bm}% bold math
\usepackage{amsmath,amsthm,amssymb}
\usepackage{graphicx}
\usepackage{epstopdf}
\usepackage{psfrag}
\def\beq{\begin{equation}}
\def\eeq{\end{equation}}

\newcommand{\affA}{Institut f{\"u}r Physik, Johannes Gutenberg-Universit{\"a}t Mainz, D-55099 Mainz, Germany}
\newcommand{\affB}{Institut f{\"u}r Quantenoptik und Quanteninformation, {\"O}sterreichische Akademie der Wissenschaften, Otto-Hittmair-Platz~1, A-6020 Innsbruck, Austria }
\newcommand{\affC}{Institut f{\"u}r Quanteninformationsverarbeitung, Universit\"at  Ulm, Albert-Einstein-Allee 11, D-89069 Ulm, Germany }
\newcommand{\affD}{Max-Planck-Institut f{\"u}r Plasmaphysik, Boltzmannstrasse 2, D-85748 Garching, Germany}
\newcommand{\affE}{Faculty of Physics, University of Warsaw, PL-00-681 Warsaw, Poland}

\usepackage{color}

\begin{document}

\preprint{}

\title{A bosonic Josephson junction controlled by a single trapped ion}

\author{R. Gerritsma}\affiliation{\affA}\affiliation{\affB}
\author{A. Negretti}\affiliation{\affC}
\author{H. Doerk}\affiliation{\affD}
\author{Z. Idziaszek}\affiliation{\affE}
\author{T. Calarco}\affiliation{\affC}
\author{F. Schmidt-Kaler}\affiliation{\affA}

\email{rene.gerritsma@uni-mainz.de}
%\affiliation{}\homepage{http://www.science.uva.nl/research/aplp/}

\date{\today}% It is always \today, today,

\begin{abstract}
We theoretically investigate the properties of a double-well bosonic Josephson junction coupled to a single trapped ion. We find that the coupling between the wells can be controlled by the internal state of the ion, which can be used for studying mesoscopic entanglement between the two systems and to measure their interaction with high precision. As a particular example we consider a single $^{87}$Rb atom and a small  Bose-Einstein condensate controlled by a single $^{171}$Yb$^+$ ion. We calculate inter-well coupling rates reaching hundreds of~Hz, while the state dependence amounts to tens of~Hz for plausible values of the currently unknown s-wave scattering length between the atom and the ion. The analysis shows that it is possible to induce either the self-trapping or the tunneling regime, depending on the internal state of the ion. This enables the generation of large scale ion-atomic wavepacket entanglement within current technology.
\end{abstract}


%\keywords{}

\maketitle

The phenomenon of tunneling in a Josephson junction is one of the most striking quantum effects that can be observed on a macroscopic scale. It has been studied in superconductors~\cite{Josephson:1962,Likharev:1979}, superfluid helium~\cite{Pereverzev:1997,Sukhatme:2001} and ultra-cold atomic gases~\cite{Albiez:2005,Gati:2007,Levy:2007,LeBlanc:2011}. In the latter, the interaction between the particles plays a crucial role, leading to phenomena such as (quantum) self-trapping~\cite{Albiez:2005}. A plethora of studies have been performed on this system
%~\cite{Millburn:1997} Either we cite all of them or we remove the reference by Milburn at this stage.
considering, for instance, one-dimensional junctions~\cite{Bouchoule:2005,Betz:2011} and analogs of photon assisted tunneling~\cite{Eckardt:2005,Grond:2011}. The bosonic Josephson junction may be well suited to study macroscopic entanglement~\cite{Bar-Gill:2011,He:2011}, like in Schr{\"o}dinger-cat-type superpositions that are instrumental in quantum information science and in the study of the crossover from quantum to classical physics. Additionally,  highly non-classical states
are an important resource to measure external (classical) fields with high precision~\cite{Riedel:2010}.

Recently, systems involving a combination of ultra-cold quantum gases and trapped ions have become available~\cite{Zipkes:2010,Schmid:2010}. These experiments have great potential to study atom-ion collisions in the quantum regime~\cite{Grier:2009,Zipkes:2010b}, cold chemistry~\cite{Rellergert:2011,Ratschbacher:2012}, and to explore sympathetic cooling of ions by means of clouds of cold atoms~\cite{Krych:2010}. Furthermore, accurately controlled trapped ions could be used to perform in situ measurements of cold atomic gases in optical lattices~\cite{Kollath:2007}, or to induce a $\mu$m-sized
bubble in a bosonic Tonks-Girardeau gas~\cite{Goold:2010}.

\begin{figure}[b!]\vspace{-0.7cm}
  \includegraphics[angle=270,width=6.5cm]{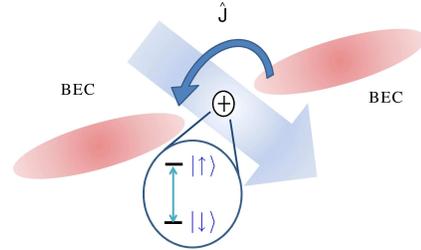}\vspace{-0.7cm}
\vspace{-0cm}  \caption{ (Color online) Bose-Einstein condensate trapped in a double well potential with an ion in the center. The internal state of the ion can be tuned by laser light and controles the tunneling rate $\hat{J}$. }
  \label{fig:setup}
\end{figure}

In this work we investigate a double well bosonic Josephson junction that is controlled by a trapped single ion as schematically shown in Fig.~1. The electric field of the ion polarizes the atoms leading to a long range attractive potential.  For small distances between the atoms and the ion, however, the interaction is strongly repulsive and depends on the relative spin orientation of the atom and ion via the singlet and triplet scattering length~\cite{Idziaszek:2011}. Hence, the tunneling is controlled by the trapped ion, especially by its internal (qubit) states $\{|\uparrow\rangle, |\downarrow\rangle\}$. This is of great interest, as it enables the generation of mesoscopic entanglement between atomic systems and ions and to study their interaction with high precision.

Our strategy to attack the hybrid Josephson junction problem consists of two steps: first, we solve the problem at the single-particle level, where only a single atom interacting with the ion is trapped in the double well potential. We will show that there is a strict relation between the inter-well coupling and the state-dependent atom-ion interaction. Moreover, we show that this dependence is rather large such that it could be observed experimentally. Then, the single-particle solutions will be used to study the many-body problem within the two-mode approximation. Finally, we shall analyze the impact of inelastic collisions and heating, and discuss experimental implementations.

We start with the single-particle problem. The first ingredient is the interaction between an atom and an ion caused by an induced atomic dipole due to the electric field of the ion. At large distances it is given by:
\begin{equation}\label{eqPot}
\lim_{r\rightarrow \infty} V_{ia}(r)=-\frac{C_4}{r^4},
\end{equation}
\noindent with $C_4=e^2\alpha_p/2$, where $e$ is the charge of the ion, and $\alpha_p$ is the static polarizability of the atom. The potential~(\ref{eqPot}) is characterised by the length scale $R^*=\sqrt{2\mu C_4/\hbar^2}$ and the energy scale $E^*=\hbar^2/(2\mu (R^*)^2)$, with $\mu$ the reduced mass. For $r\rightarrow 0$, Eq.~(\ref{eqPot}) does not hold anymore and the potential becomes strongly repulsive. The exact form of the potential in this regime is generally not known. Given this, we will use quantum defect theory, where all the information on the short-range interaction is put in a single quantum defect parameter, the so-called short-range phase $\phi$, which is related to the ion-atom s-wave scattering length $a_{ia}=-R^*\cot(\phi)$~\cite{Idziaszek:2011,Doerk:2010}.

\begin{figure}[t]
  \includegraphics[width=8.0cm]{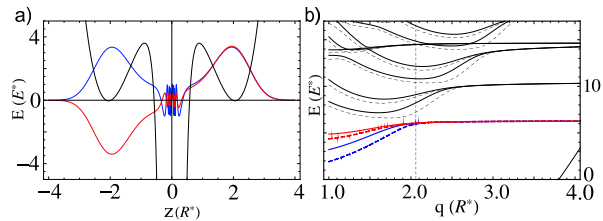}\vspace{-0.2cm}
  \caption{(Color online) a)  Wave functions for the ground states $\Phi_g(z)$ (blue) and $\Phi_e(z)$ (red), at $q=2.06R^*$; the black curve shows the potential with $\omega_q=2\pi$~3.9~kHz. The wavefunctions are  multiplied by a constant for clarity. b) Spectrum for a double well as a function of $q$ and short-range phases of $\phi=-\pi/4$ (solid lines) and $\phi=-\pi/3$ (dashed lines) with the ground states shown in bold red and blue. The dashed vertical line marks the separation $q=2.06R^*$.}
  \label{fig:specs}
\end{figure}

\begin{figure}[t]
  \includegraphics[width=7.0cm]{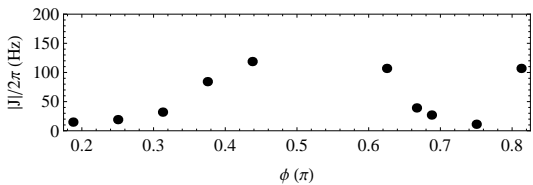}\vspace{-0.2cm}
  \caption{ Inter-well coupling $J$ for various short-range phases with $\omega_q=2\pi$~3.9~kHz and $q=2.06R^*$.  }
  \label{fig:omvsphi}
\end{figure}\vspace{-0.1cm}

Now we consider the experimentally realistic case in which the ion is trapped much more tightly than the atom, $\omega_i \gg \omega_a$, with $\omega_{i,a}$ the trapping frequency of the ion and atom. We also assume that the ion motion is cooled to the ground state (e.g., by sideband cooling). These assumptions enable us to make an approximation in which the ion is treated statically at the position $\mathbf{r}_i=0$. We further assume that the ion and atom cannot undergo spin changing collisions, allowing the use of a single channel model. This is justified when the singlet and triplet scattering lengths between the ion and atom are similar~\cite{Doerk:2010}, or when there is only one open channel for the considered states. Under these conditions the Schr{\"o}dinger equation for the atomic wave function, in units of $R^*$ and $E^*$, is given by:

\beq\label{eqspherical}
E\bm{\psi}(\mathbf{r})=\left(-\frac{\mu\nabla^2}{m_a}+V_{dw}(\mathbf{r})-\frac{1}{r^4}\right)\bm{\psi}(\mathbf{r}),
\eeq
\noindent where $V_{dw}(\mathbf{r})$ is the atomic double well potential, and $m_a$ the atomic mass. We consider a double well of the form:

\beq\label{eqVdimless}
V_{dw}(\mathbf{r})=\frac{b}{E^*q^4}\left(z^2-q^2\right)^2+\alpha \frac{\mu}{m_a} (x^2+y^2),
%V_{dw}(\mathbf{r})=\left(r^2+\frac{q^2}{4}-\frac{3}{2} z^2+\frac{1}{4q^2}z^4\right)\frac{\alpha \mu}{m_a},
\eeq

%\change{which has minima at $\mathbf{r}=(0,0,\pm q)$ with local trapping frequency  $\omega_q=\omega_a$, with $\omega_a$ the atomic trapping frequency in the transverse $(x,y)$-directions.}{

\noindent which has minima at $\mathbf{r}=(0,0,\pm q)$ with local trapping frequency  $\omega_q=\sqrt{\frac{8b}{m_aq^2}}$, and  $b$ is the barrier height between the wells.  For convenience we assume that $\omega_q=\omega_a$, with $\omega_a$ the atomic trapping frequency in the  $(x,y)$-directions. We consider varying the inter-well separation $q$ without changing the local trapping parameters: only the inter-well barrier height is changed. Besides this, $\alpha = (R^*/l_0)^4$ and $l_0=\sqrt{\hbar/m_a \omega_a}$. 

For $r\rightarrow 0$ the atomic trapping potential and energy are negligible compared to the atom-ion interaction and the solution can be obtained analytically: $\tilde{\psi}_l(r)=\tilde{N} \sqrt{r} \left[J_{l+1/2}(\xi)+\tan(\delta)Y_{l+1/2}(\xi)\right]$. We use this solution as a boundary condition for numerically solving Eq.~(\ref{eqspherical})~\cite{Supp_mat}. Here, $J_{l+1/2}(\xi)$ and $Y_{l+1/2}(\xi)$ are spherical Bessel functions with orbital angular momentum quantum number $l$, $\xi=\sqrt{m_a/\mu r}$ and $\tilde{N}$ is a normalization constant.  The mixing angle $\delta$ is related to the short-range phase: $\delta=-\phi-l\pi/2$. In our study we neglect small corrections due to the $l$-dependence of $\phi$, because in the ultracold collisional regime basically only s-wave scattering occurs~\cite{Idziaszek:2011}. 

As a specific example to our model, we consider a single trapped $^{87}$Rb atom in the $|F_a=2,m_{Fa}=2\rangle$ state and a $^{171}$Yb$^+$ ion with logical states $|\downarrow\rangle \equiv |F_i=0,m_{Fi}=0\rangle$ and $|\uparrow\rangle \equiv |F_i=1,m_{Fi}=1\rangle$. When the ion is in state $|\uparrow\rangle$, the dynamics is described by the triplet scattering length $a_t$. When the ion is in the state $|\downarrow\rangle$ the interaction depends on both the singlet and triplet scattering length. Since these values are not currently known, we rather assume plausible values for the corresponding short-range phases $\phi_{\uparrow}$ and $\phi_{\downarrow}$. 

We set the parameter $\alpha$ to 10, corresponding to $\omega_q =2\pi$~3.9~kHz. For the chosen atom and ion $R^*$=0.306~$\mu$m and $E^*/\hbar=2\pi$~935~Hz. In Fig.~\ref{fig:specs} the spectrum as a function of the inter-well separation $q$ is shown. For large $q$, the spectrum is equivalent to that of a harmonic oscillator, since there is no coupling between the wells. For smaller $q$ the levels split in two as the inter-well coupling increases. The ground state splits in the symmetric (ground) and anti-symmetric (excited) superpositions $\Phi_{g,e}(\mathbf{r})=(\Phi_L (\mathbf{r})\pm \Phi_R(\mathbf{r}))/\sqrt{2}$ with energies $E_g$ and $E_e$ as shown in Fig.~\ref{fig:specs}. Here, $\Phi_{R,L} (\mathbf{r})$ denotes the left/right localized wavepackets. When the splitting $\Delta E=E_e-E_g$ is much smaller than the energy gap to the other levels in the spectrum, a two-mode approximation  can be employed, with the inter-well coupling given by $J=\Delta E/(2\hbar)$. As Fig.~\ref{fig:specs}b) shows, this coupling depends on the internal state of the ion. In this example we chose $\phi_{\downarrow}=-\pi/3$ and $\phi_{\uparrow}=-\pi/4$. At the position $q=2.06 R^*$, the coupling amounts to $J_{\uparrow}=2\pi~11.2$~Hz  and $J_{\downarrow}=2\pi~39.2$~Hz. Without an ion the tunneling rate would be only~$2\pi~0.96$~Hz. In Fig.~\ref{fig:omvsphi}, the inter-well coupling for various other short-range phases $\phi$ is shown. This analysis clearly shows that there is a relatively large dependence of the inter-well coupling on the short-range phase, which should be experimentally observable, and $J$ depends on the qubit state via $\phi$. Hence, we can treat the inter-well coupling as an operator and write it as: $\hat{J}=J_{\downarrow}|\downarrow\rangle \langle \downarrow|+J_{\uparrow}|\uparrow\rangle \langle \uparrow|$. Moreover, typical spectra show much larger energy gaps to neighbouring states, such that the two-mode approximation is well justified in a many-body calculation.

\begin{figure}[t!]
 \includegraphics[width=8.5cm]{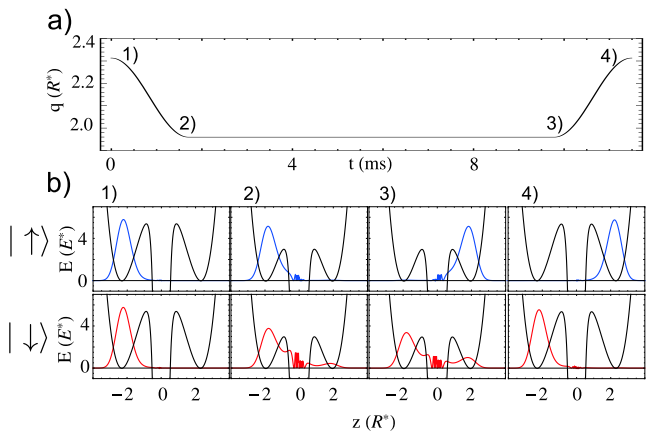}
  \caption{(Color online) a) Time dependence of the inter-well distance $q$ for generating state dependent tunneling. b) Probability distributions at various times. The initial state prepared at $q=2.3R^*$ is an equal superposition of the two ground states $\psi(z,t=0)=(\Phi_g(z)-\Phi_e(z))/\sqrt{2}$. At the end of the sequence the atom has tunneled when the ion is prepared in $|\uparrow \rangle$ (top panel b) and returned to the left well after tunneling multiple times when the ion is prepared in $|\downarrow \rangle$ (lowest panel). For the calculation we assume $\phi_{\uparrow}=-\pi/4$ and $\phi_{\downarrow}=-\pi/3$. At the end of the sequence $>98$\% is in the right/left well respectively. }
  \label{fig:evols}
\end{figure}

We can use the state dependence in the double-well to design an experimental sequence that entangles the internal state of the ion with the position of the atom. To this aim, we introduce time dependence into the well separation $q(t)$ [Fig.~\ref{fig:evols} a)] and solve the time-dependent Schr\"odinger equation numerically for the sequence: 1) The ion is prepared in the internal state $(|\downarrow\rangle+|\uparrow\rangle)/\sqrt{2}$ and the atom is initially trapped in the left well. Then we adiabatically bring the wells together. 2)  The atom tunnels and we wait until the atom is either in the right or left well, depending on the (internal) state of the ion. 3) We take apart the wells again such that tunneling stops. 4) The resulting atom-ion state is the entangled state $\vert\psi_{final}\rangle\propto |\downarrow\rangle |\Phi_R\rangle+e^{i\beta}|\uparrow\rangle |\Phi_L\rangle$, where the phase $\beta$ depends on the details of the sequence. A detailed calculation of this process is shown in Fig.~\ref{fig:evols}b), where the potential and the atomic probability distributions are plotted. The setup can also be thought of as an interferometer, in which the entangling-disentangling dynamics can be monitored to study the interaction between the ion and the atom. Similar studies have been performed e.g. with cold Rb atoms in optical lattices~\cite{Widera:2009}, where differences in scattering length for different spin states lead to such dynamics. 

Now let us consider the many-particle scenario. To this aim we apply a two-mode approximation in which the quantum field operator is expanded in the local modes $\Phi_{L,R}(\mathbf{r})$ of the double well potential, that is, 
$\hat\psi(\mathbf{r},t)=\Phi_L(\mathbf{r})\hat c_L+\Phi_R(\mathbf{r})\hat c_R$, where $\hat{c}_i$ ($\hat{c}^{\dag}_i$) with $i=L,R$ is the annihilation (creation) 
operator of an atom in the left or right well.
The local modes are the single-particle ground states of the system that we have obtained already. The two-mode approximation provides a reasonable description of the dynamics when many-body interactions yield only slight modifications to the ground state. This requires $N \ll \bar{l}_0/a_a$, with $N$ the number of atoms, $\bar{l}_0=(l_xl_yl_z)^{1/3}$ the size of the ground state wave packet, and $a_a$ the atomic s-wave scattering length. Under these assumptions, the system can be effectively described by the following Bose-Hubbard (BH) Hamiltonian~\cite{Gati:2007,Millburn:1997}:
\beq\label{eq_HBJJ}
\hat{H}_{BH} = \hbar\hat{J} \hat\alpha+\hbar\hat{U} \hat n^2,
\eeq
where $\hat\alpha = \hat{c}_L^{\dag} \hat{c}_R+\hat{c}_R^{\dag}\hat{c}_L$ is the tunneling operator, and 
$\hat n = (\hat{c}_R^{\dag} \hat{c}_R-\hat{c}_L^{\dag}\hat{c}_L)/2$ describes the population imbalance between the two wells. Because of the spin-dependence of the atom-ion interaction we also write the on-site atom-atom interaction as an operator $\hat{U}$.

%=U_{\downarrow}|\downarrow\rangle \langle \downarrow|+U_{\uparrow}|\uparrow\rangle \langle \uparrow|$.

We recall that the above outlined two-mode model enables to identify, depending on the ratio $U/J$, three different regimes~\cite{Gati:2007}: 
the``Rabi'' regime (i.e., weak atom-atom interaction); the intermediate 
``Josephson'' regime, and the ``Fock'' regime (i.e., strong atom-atom interaction). In our study we are interested in the Rabi and Fock regimes. We show that
depending on the internal state of the ion, it is possible to induce either ``Rabi'' oscillations between the two wells or to keep the condensate trapped 
in a single well (i.e., the self-trapping regime). It is useful to introduce the interaction parameter $\Lambda = UN/2J$. 
Self-trapping regime is predicted to occur for $\Lambda > \Lambda_c = 2$~\cite{Sakmann:2009}.

Now the on-site interaction energy can be estimated by using the single-particle wave functions leading to $U=E^*\frac{U_0}{\hbar}\int d \mathbf{r} |\Phi_L(\mathbf{r})|^4$ with $U_0=8\pi a_a\mu/m_a R^*$~\cite{Millburn:1997}. 
Here we neglected terms such as $E^*\frac{U_0}{\hbar}\int d \mathbf{r} |\Phi_L(\mathbf{r})|^2|\Phi_R(\mathbf{r})|^2$, which are much smaller than $U$. 
Then, in order to quantify the quality of the description in localized wavepackets, which is crucial for the BH model, we calculate the probability $\epsilon$ of 
finding the atoms in the 'wrong' region, namely  $\epsilon=\int_R |\Phi_L (\mathbf{r})|^2d\mathbf{r}^3$, and the integral taken over $R$, the right half space, ($z>0$).

\begin{figure}[t!]
 \includegraphics[width=7cm]{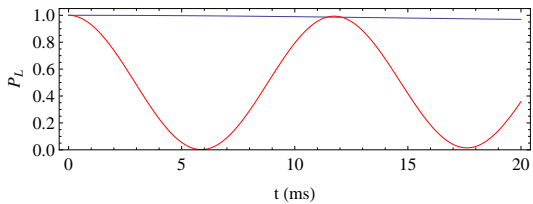}
  \caption{(Color online)  Relative population in the left well as a function of time for $|\uparrow\rangle$ (blue) and $|\downarrow\rangle$ (red)  $N=20$ atoms. Obtained by integrating Eq.~(\ref{eq_HBJJ}) numerically, assuming $\phi_{\uparrow}=-\pi/4$ and $\phi_{\downarrow}=\pi/4$. }
  \label{fig:becdynamics}
\end{figure}

Let us consider a weaker trap, such that both the Rabi and the self-trapping regimes are achieved for a relatively large number of atoms. By choosing $\omega_a = \omega_q = 2\pi$~200~Hz ($\alpha=0.026$) 
and $q=6.39$~$R^*$, the two-mode approximation requires $N \ll 100$ and we find the following  parameters: $U_{\uparrow}=2\pi$~0.9~Hz, 
$J_{\uparrow}=2\pi$~1.7~Hz for $\phi_{\uparrow}=-\pi/4$ ($a_{ia}=R^*$) with $\epsilon = 0.004$. Whereas for negative scattering lengths $a_{ia}=-R^*$, 
i.e., $\phi_{\downarrow}=\pi/4$, we have $U_{\downarrow}=2\pi$~1.0~Hz, $J_{\downarrow}=2\pi$~42.7~Hz  with $\epsilon = 0.017$. For $N=20$ atoms 
we have $\Lambda_c^{\uparrow}=5.2$ (i.e., self-trapping regime) and $\Lambda_c^{\downarrow}=0.23$ (i.e., the Rabi regime). The resulting BH dynamics 
is shown in Fig.~\ref{fig:becdynamics}, which clearly shows that large scale entanglement can be achieved. For $N=100$ atoms, we would find 
$\Lambda_c^{\uparrow}=25.8$ and $\Lambda_c^{\downarrow}=1.17$. For $\phi_{\uparrow}=\pi/3$, i.e. ($a_{ia}=-R^*/\sqrt{3}$), we have for $N=20$ atoms 
$\Lambda_c^{\uparrow}=2$ and for $N=100$ atoms $\Lambda_c^{\uparrow}=9.8$.

We underscore, however, that in the self-trapping regime the BH-model is in disagreement with the full numerical many-body calculation especially at long time scales~\cite{Sakmann:2009}. The time scales are set by the Rabi period $t_{Rabi}=\pi/J$. With the above outlined parameters we find: $t_{Rabi}^{\uparrow}/t_{Rabi}^{\downarrow}= 25$~(8) for $\phi_{\uparrow}=-\pi/4$~$(\pi/3)$. Therefore, we might expect reasonable agreement between the Bose-Hubbard model and a full quantum many body calculation at short times, that is, $t\ll t_{Rabi}^{\uparrow}$ but $t\sim t_{Rabi}^{\downarrow}$ and not too far into the self-trapping regime, as in Fig.~\ref{fig:becdynamics}.

We will now discuss some of the limitations of our model and possible experimental issues. As we already noted before, the atom and ion could change their spin states in a collision~\cite{Idziaszek:2011}. When the ion is in the state $|\uparrow\rangle$, spin changing collisions cannot happen due to the conservation of the quantum number $M_F = m_a+m_i=3$: there are no other states with $M_F=3$. When the ion is in the state $|\downarrow\rangle$, $M_F=2$, we have to consider the mixing with the states $|F_i,m_{Fi},F_a,m_{Fa}\rangle=|1111\rangle$, $|1121\rangle$ and $|1022\rangle$. However, all these states have significantly higher dissociation energies, since the hyperfine splitting of $^{171}$Yb$^+$ (12.6~GHz) is much larger than that of $^{87}$Rb (6.8~GHz). Thus, only resonances with bound states can occur, which have a much lower density than trap states. Consequently, we can reasonably assume that also for $|0022\rangle$ no spin flips occur. 

Concerning inelastic collisions, it is important to note that our proposal requires low atomic densities. We estimate the peak density of the atomic wavepackets to be about $10^{20}$/m$^3$. Now, assuming that many-body interactions do not perturb significantly the form of the wavefunction, this would lead to loss rates of a few Hz~\cite{Zipkes:2010}. Secondly, in our proposal the atoms and ion are not directly overlapped, the ion rather perturbs the atomic wavefunction while it is held in its own trapping potential, reducing the density and terefore the inelastic scattering a factor $>$10 more.  This setup may also reduce heating due to the 'micromotion' of the ion in its time-dependent trap~\cite{Nguyen:2012,Cetina:2012}. However, even in the presence of moderate inelastic collision rates, our proposal could be implemented, since these effects are easily separated from experimental errors: one would simply observe that there is no (Yb$^+$) ion at the end of the experiment. 

Regarding cooling and heating mechanisms, we first note that the ion can be cooled to the ground state with effectively unit efficiency using sideband coolings. However, given the time scales required for the proposed experiments, heating may cause problems. In state of the art ion traps, such as in~\cite{Kirchmair:2009}, heating rates of a few phonons/s are achieved, such that an experiment on the tens of ms timescale should be possible. The coherence of the atomic wavepacket can be sustained for tens of ms even for well-separations of several $\mu$m~\cite{Albiez:2005}.

Finally we mention that Josephson junctions for cold atoms have been implemented in optical dipole traps and in micro-fabricated magnetic traps~\cite{Albiez:2005,Gati:2007,Levy:2007,LeBlanc:2011}. Such systems can be combined with ion trap technology. One interesting approach would be to integrate atom and ion traps in a single micro-trap. Recent experiments have demonstrated micro fabricated wires or permanent magnets on ion surface traps~\cite{Leibfried:2007,Wang:2009,Ospelkaus:2011,Welzel:2011}, with the purpose of performing quantum logic gates without lasers by magnetic coupling to the ions. In Refs.~\cite{Wang:2009,Welzel:2011} magnetic field gradients of 2-20~T/m are reported which would make it possible to magnetically trap $^{87}$Rb atoms with trapping frequencies reaching 20~kHz, 50~$\mu$m above the trap surface. 

In conclusion, we studied the tunneling dynamics of a single atom and a small BEC in combination with a trapped ion. We found that the junction transmission can be controlled by the internal state of the ion making it possible to engineer large ion-matterwave entangled states. The system could be scaled up by considering Josephson junctions coupled by strings of ions. The ion-atom hybrid quantum system may lead to precision measurement of external fields, or collisional properties where short-range interactions may be measured by determining tunneling frequencies.

\begin{acknowledgements}
This work was supported by the EU project AQUTE (F.S., T.C.), QIBEC and PICC (T.C.), the Marie-Curie Programme of the EU through Proposal Nr. 236073 (OPTIQUOS) within the 7th European Community Framework Programme (A.N.), and the DFG within the Grant No. SFB/TRR21 (T.C.).
\end{acknowledgements}

\section{Additional material}

\subsection{Diagonalisation of the double well hamiltonian.}

To numerically solve Eq.~(2) in the paper, we make use of renormalized Numerov method~\cite{Johnson:1977} with 

\beq
\tilde{\psi}_l(r)=\tilde{N} \sqrt{r} \left[J_{l+1/2}(\xi)+\tan(\delta)Y_{l+1/2}(\xi)\right]
\eeq
as a boundary condition. Our approach is to obtain a set of basis functions $\bm{\psi}^0(\mathbf{r})$ with energies  $E^0$ in which to expand the solutions to the double-well problem. The superscript $0$ is put to mean that there is no inter-well separation, as will become clear below. A convenient choice for this purpose is an isotropic atomic harmonic oscillator $V(\mathbf{r})=\alpha \mathbf{r}^2 \mu/m_a$ with a single ion trapped in its center, where $\alpha = (R^*/l_0)^4$ and $l_0=\sqrt{\hbar/m_a \omega_a}$ is the size of the atomic ground state. The solutions $\bm{\psi}^0$ have the form $\bm{\psi}^0=Y_l^m \psi_{nl}(r)/r$, where $Y_l^m$ are spherical harmonics and $\psi_{nl}(r)/r$ denotes the radial wavefunctions to be found. We obtain a set of 1250 eigenfunctions and energies including bound as well as trap states~\cite{Doerk:2010}. The procedure is similar to the one used in Ref.~\cite{Doerk:2010}. 

Now we consider the double well of the form:

\begin{eqnarray}\label{eqVdimless}
V_{dw}(\mathbf{r})&=&\frac{b}{E^*q^4}\left(z^2-q^2\right)^2+\alpha \frac{\mu}{m_a} (x^2+y^2)\nonumber\\
&\stackrel{\omega_q=\omega_a}{=}&\left(r^2+\frac{q^2}{4}-\frac{3}{2} z^2+\frac{1}{4q^2}z^4\right)\frac{\alpha \mu}{m_a}.\nonumber
\end{eqnarray}
The first term in Eq.~(\ref{eqVdimless}) corresponds to the harmonic potential used for
determining the basis functions $\psi_{nl}^0(\mathbf{r})$, and therefore this part of the Hamiltonian is diagonal. On the other hand, for the remaining terms in Eq.~(\ref{eqVdimless}), $\psi_{nl}^0(\mathbf{r})$ are not eigenfunctions, and thereby we have to compute the corresponding matrix elements (using $z = r \cos\theta$). We use the expansion onto the solutions of the isotropic harmonic well $\bm{\psi}(\mathbf{r})=\sum_{k} c_{k} \psi^0_{k}(\mathbf{r})$, where $k$ denotes the pair of quantum numbers $(n,l)$. Since the potential does not depend on the azimuthal direction, the quantum number $m$ is conserved and, without loss of generality, we set $m=0$. Both the wavefunctions and the corresponding energies are obtained by diagonalising the Hamiltonian, whose matrix elements are given by:
\begin{eqnarray}
H_{kk'}&=&\left(E^0_{k}+\frac{\mu\alpha q^2}{4m_a}\right)\delta_{kk'}-\frac{3\mu\alpha}{2 m_a} M^{(2)}_{kk'}+\frac{\mu\alpha }{4m_aq^2} M^{(4)}_{kk'},\nonumber\\
M^{(j)}_{kk'}&=&\int d\mathbf{r}^3 \psi^{0*}_{k'}(\mathbf{r}) \,\cos^j(\theta)\, r^{j} \,\psi^{0}_k(\mathbf{r}),
\end{eqnarray}

\noindent with $j=2,4$. The clear advantage of this procedure is that the matrix elements do not depend on $q$, such that spectra are easily obtained as a function of the inter-well separation.

\subsection{Short range state dependence}

For $r\rightarrow 0$ the interaction between atom and ion is described by the quantum numbers $|FM_FIS\rangle$~\cite{Idziaszek:2011}, with $\mathbf{F}=\mathbf{F}_a+\mathbf{F}_i$, the total nuclear spin $\mathbf{I}=\mathbf{i}_a+\mathbf{i}_i$ and total electron spin $\mathbf{S}=\mathbf{s}_a+\mathbf{s}_i$. Depending on whether $S=0$ or $S=1$, the interaction is described by the singlet or triplet scattering length $a_s$ and $a_t$, respectively. The quantum states for large $r$, labeled by the hyperfine states $|F_i,m_{Fi},F_a,m_{Fa}\rangle$ are connected to the quantum numbers for the short-range interaction via a frame transformation~\cite{Idziaszek:2011}. When the ion is in state $|\uparrow\rangle$, the dynamics is described by the triplet scattering length $a_t$, since $S=1$.  On the other hand, when the ion is in the state $|\downarrow\rangle$, the dynamics depends both on the singlet and triplet scattering length, since this state is a linear combination of short-range states $|FM_FIS\rangle$, in particular we have: 

\begin{equation}
|F_a=2,m_{F_a}=2\rangle|\downarrow\rangle=-\frac{1}{2}|2220\rangle+\sqrt{\frac{3}{8}}|2211\rangle+\sqrt{\frac{3}{8}}|2221\rangle\nonumber
\end{equation}

\noindent which can be obtained by construction of the frame transformation in terms of Clebsch-Gordan coefficients and Wigner 9-j symbols (see Ref.~\cite{Idziaszek:2011}). Here, the right hand side is labeled according to $|FM_FIS\rangle$.

%%-------------------------------
%\bibliography{biblio-RG,amo}
%-------------------------------

\end{document}